\documentstyle [12pt] {article}
\makeatletter

\@addtoreset{equation}{section}

\ifcase \@ptsize
\or 
\or 
\fi
\advance\textheight by \topskip

\ifcase \@ptsize
\evensidemargin
\or 
\or 
\fi

\def\@citex[#1]#2{%
\if@filesw \immediate \write \@auxout {\string \citation {#2}}\fi
\@tempcntb\m@ne \let\@h@ld\relax \def\@citea{}%
\@cite{%
  \@for \@citeb:=#2\do {%
    \@ifundefined {b@\@citeb}%
      {\@h@ld\@citea\@tempcntb\m@ne{\bf ?}%
      \@warning {Citation `\@citeb ' on page \thepage \space undefined}}%
      {\@tempcnta\@tempcntb \advance\@tempcnta\@ne%
      \@tempcntb\number\csname b@\@citeb \endcsname \relax%
      \ifnum\@tempcnta=\@tempcntb 
        \ifx\@h@ld\relax%
          \edef \@h@ld{\@citea\csname b@\@citeb\endcsname}%
        \else%
          \edef\@h@ld{\ifmmode{-}\else--\fi\csname b@\@citeb\endcsname}%
        \fi%
      \else
        \@h@ld\@citea\csname b@\@citeb \endcsname%
        \let\@h@ld\relax%
      \fi}%
    \def\@citea{,\penalty\@highpenalty\,}%
  }\@h@ld%
}{#1}}

\makeatother
\begin{document}
\hfuzz=100pt
\textheight 24.0cm
\topmargin -0.5in
%
%
%
%
\newcommand{\be}{\begin{equation}}
\newcommand{\ee}{\end{equation}}
\newcommand{\bea}{\begin{eqnarray}}
\newcommand{\eea}{\end{eqnarray}}
\begin{titlepage}
\makeatletter
\def \thefootnote {\fnsymbol {footnote}} \def \@makefnmark {
\hbox to 0pt{$^{\@thefnmark }$\hss }}
\makeatother
\begin{flushright}
BONN-TH-94-03\\
April, 1994\\
hep-th/xxxxxxx
\end{flushright}
\vspace{1.5cm}
\begin{center}
{ \large \bf Wess-Zumino-Novikov-Witten Models Based on Lie Superalgebras}\\
\vspace{2cm}
{\large\bf Noureddine Mohammedi}
\footnote{e-mail: nouri@avzw02.physik.uni-bonn.de}
\footnote
{Work supported by the Alexander von Humboldt-Stiftung.}
\\
\vspace{.5cm}
Physikalisches Institut\\
der Universit\"at Bonn\\
Nussallee 12\\ D-53115 Bonn, Germany\\

\baselineskip 18pt
\vspace{1cm}
{\large\bf Abstract}
\end{center}
The affine current algebra for Lie superalgebras is examined.
The bilinear invariant forms of the Lie superalgebra can be
either degenerate or non-degenerate. We give the conditions for a
Virasoro construction, in which the currents are primary fields of
weight one, to exist. In certain cases, the Virasoro central charge
is an integer equal to the super dimension of the group supermanifold.
A Wess-Zumino-Novikov-Witten action based on these Lie superalgebras is
also found.  \\
\setcounter {footnote}{0}
\end{titlepage}
\baselineskip 20pt
\section{Introduction}
In this note we explore from a mathematical point of view
a class of exact conformal field theories based
on Wess-Zumino-Novikov-Witten (WZNW) built on  Lie superalgebras.
The group supermanifold corresponding to these Lie
superalgebras could be neither compact nor semi-simple.
In these models the Virasoro central charge does, in
general, depend on the level of the affine Lie superalgebra.
However, for certain cases the central charge is equal to
the super dimension (the dimension of the bosonic
part of the Lie superalgebra minus the dimension
of the fermionic part) of the  group supermanifold.
\par
The motivation behind this study stems from the recent
spate of interest in constructing WZNW models
based on non-semi-simple groups [1--8]. This is
because these models lead to exact string backgrounds
having a target space dimension equal to the integer
Virasoro central charge of the affine non-semi-simple
algebra [3,4,5,6].
The first of these models, constructed by Nappi and Witten [1],
was based on a central extension of the two-dimensional
euclidean group and describes a string propagating in
a homogeneous four-dimensional spacetime in the background
of a gravitational plane wave. This construction of WZNW
models for non-semi-simple groups was subsequently
extended to other groups and generalised to higher
dimensions [3,4,7,8]. {}Furthermore, all these models describe
the propagation of strings on a target space whose metric
possess a covariantly null Killing vector [9--14].
\par
In ordinary WZNW models the invertibility of the Cartan-Killing
invariant bilinear form is crucial for the Sugawara construction
of the corresponding stress tensor. But since for non-semi-simple
groups this bilinear form is degenerate, the Sugawara construction
(that is a  Virasoro construction in which the currents are
primary fields of weight one) does not
exist. However, for all the  non-semi-simple algebras
studied so far [1,3,4,6,7] it has been possible to
find non-degenerate
invariant bilinear forms. These invertible bilinear forms allow for
a Sugawara-type
construction leading to integer Virasoro central charges.
\par
In [5] we gave the conditions for an affine Lie algebra (semi-simple
or non-semi-simple) to admit a Sugawara-type construction with
respect to which the currents are primary fields of conformal dimension
one. An expression for the Virasoro central charge was also found and
from which one sees that this central charge is not necessarily
an integer number. A search, carried out in [6], for
non-semi-simple Sugawara-type
construction having non-integral value of the central charge
shows that this construction necessarily factorises into
a semi-simple standard Sugawara construction and a
non-semi-simple one with integral central charge.
\par
In this paper we extend the analyses and constructions of ref.[5] to
current algebras based on  Lie superalgebras.  In an affine
Lie superalgebra one encounters two invariant bilinear forms; one
symmetric and the other antisymmetric. In the literature
super Sugawara constructions corresponding to affine Lie
superalgebras were extensively studied in [15]. The existence of these
super Sugawara construction relies on the invertibility of the
two bilinear forms entering the affine  Lie superalgebra. The aim
of this paper is to explore the possibility of finding super
Sugawara-type constructions even when the two bilinear forms
of the affine Lie superalgebra are degenerate. In other words,
using other invertible bilinear forms, we construct a solution
of the super master equation [16] with the additional property
that all the currents are primary field of conformal dimension
one.
\par
The paper is organised as follows: In section two we review the  Lie
supercurrent algebra and construct the energy-momentum tensor.
The conditions under which the generators of the  Lie supercurrent
algebra are primary fields of conformal dimension one, are spelled out.
Section three is dedicated to the construction of a WZWN action based
on  Lie superalgebras.

\section{The Lie Supercurrent Algebra}

A Lie superalgebra, ${\cal G}$, is expressed in terms of a
set of bosonic generators, $t^a\in {\cal G}_0$, together with a set of
fermionic
generators, $S^\alpha\in {\cal G}_1$.
Its affine extension (for which $t^a$ and $S^\alpha$ are the zero modes) is
given by the commutation relations [15]
\bea
\left [t^a_m\,,\,t^b_n\right ]&=&f^{ab}_ct^c_{m+n} + m g^{ab}\delta_{m+n,0}
\nonumber\\
\left [t^a_m\,,\,S^\alpha_r\right ]&=&N^{a\alpha}_\beta
S^\beta_{m+r}
\nonumber\\
\left\{S^\alpha_r\,,\,S^\beta_s\right\}&=&R^{\alpha\beta}_at^a_{r+s}
+r\varphi^{\alpha\beta}\delta_{r+s,0}\,\,\,.
\eea
Here $g^{ab}$ and $\varphi^{\alpha\beta}$ are two invariant bilinear forms
of the Lie superalgebra and they are, respectively, symmetric and
antisymmetric.
We do not assume these two bilinear forms to be invertible.
\par
The structure constants of the above algebra are related by the
super Jacobi identities. {}Firstly, we have the usual relations that one
encounters in bosonic Lie algebras, namely
\bea
&f^{ab}_cf^{cd}_e + f^{da}_cf^{cb}_e +f^{bd}_cf^{ca}_e =0&
\nonumber\\
&g^{ab}f^{cd}_{b}+g^{cb}f^{ad}_{b}=0\,\,\,.&
\eea
Secondly, the symplectic form $\varphi^{\alpha\beta}$ satisfies
\be
N^{a\alpha}_\beta\varphi^{\beta\gamma}
-N^{a\gamma}_\beta\varphi^{\beta\alpha}=0\,\,\,.
\ee
Thirdly, the structure constants $N^{a\alpha}_\beta$
and $R^{\alpha\beta}_a$ have to fulfill
\bea
&N^{a\alpha}_\gamma R^{\gamma\beta}_b +
N^{a\beta}_\gamma R^{\gamma\alpha}_b=f^{ac}_bR^{\alpha\beta}_c&
\nonumber\\
&N^{a\alpha}_\tau N^{b\tau}_\beta-
N^{b\alpha}_\tau N^{a\tau}_\beta=-f^{ab}_cN^{c\alpha}_\beta
\,\,\,.&
\eea
The last equation means that the matices $(-N^a)^\alpha_\beta$ define
a representation
of the bosonic part of the  Lie superalgebra. The structure constants
$R^{\alpha\beta}_a$ are related to $N^{a\alpha}_\beta$ by the relation
\be
R^{\alpha\beta}_ag^{ab}- N^{b\alpha}_\tau\varphi^{\tau\beta}=0\,\,\,.
\ee
Notice that $R^{\alpha\beta}_a$ is indeed symmetric in $\alpha$
and $\beta$. {}Finally, we have
\be
R^{\alpha\beta}_aN^{a\gamma}_\tau+
R^{\gamma\alpha}_aN^{a\beta}_\tau+
R^{\beta\gamma}_aN^{a\alpha}_\tau=0\,\,\,.
\ee
\par
In terms of operator product expansions, the above commutation relations are
equivalent to\footnote{To get the usual level $k$, one simply scales
$g^{ab}$ and $\varphi^{\alpha\beta}$ by a factor $k$.}
\bea
J^a(z)J^b(w)&=&{g^{ab}\over {(z-w)^2}}+f^{ab}_{c}
{J^c(w)\over{(z-w)}}
\nonumber\\
J^a(z)S^\alpha(w)&=&N^{a\alpha}_{\tau}
{S^\tau (w)\over{(z-w)}}
\nonumber\\
S^\alpha (z)S^\beta (w)&=&{\varphi^{\alpha\beta}\over
{(z-w)^2}}+R^{\alpha\beta}_{c}
{J^c(w)\over{(z-w)}}\,\,\,.
\eea
Here $J^a(z)=\sum_n t^a_nz^{-n-1}$ are the bosonic currents of the even
part of the Lie superalgebra, and $S^{\alpha}(z)=\sum_r S^\alpha_r z^{-r-1}$
constitute the fermionic currents of the odd part of the
Lie superalgebra. The label $n$ is an integer while  $r$ can be
either integer or half-integer.
\par
Let us assume now that we have two bilinear forms
$\Omega^{ab}$ and $\psi^{\alpha\beta}$ which are respectively
symmetric and antisymmetric and invertible, namely
$\Omega^{ab}\Omega_{bc}=\delta^a_c$ and
$\psi^{\alpha\beta}\psi_{\beta\tau}=\delta^\alpha_\tau$. These two
invertible bilinear forms are also assumed to satisfy
\bea
&\Omega^{ab}f^{cd}_{b}+\Omega^{cb}f^{ad}_{b}=0&
\nonumber\\
&N^{a\alpha}_\beta\psi^{\beta\gamma}
-N^{a\gamma}_\beta\psi^{\beta\alpha}=0&
\nonumber\\
&R^{\alpha\beta}_a\Omega^{ab}- N^{b\alpha}_\tau\psi^{\tau\beta}=0\,\,\,.&
\eea
It is then easy to verify that the Casimir operator
of the Lie superalgebra (the algebra of the zero modes
$t^a$ and $S^\alpha$ in (2.1)) is given by
\be
Q=\Omega_{ab}t^at^b+\psi_{\alpha\beta}S^\alpha S^\beta\,\,\,.
\ee
\par
The energy-momentum tensor is taken to be given by the following
quadratic combination of bosonic and fermionic currents [16]
\be
T(z)= C_{ab}:J^aJ^b:(z) + D_{\alpha\beta}:S^{\alpha}S^\beta:(z)\,\,\,,
\ee
where $C_{ab}$ is symmetric and $D_{\alpha\beta}$ is antisymmetric.
We would like now to determine these last two tensors by requiring
that both $J^a(z)$ and $S^\alpha (z)$ are primary fields
of conformal dimension equal to one with respect to the
above energy-momentum tensor.
\par
In order to perform the different
operator product expansions, we define
the normal ordered product of two operators $A(z)$ and $B(z)$
at coincident points as [17]
\be
:AB:(z)={1\over 2\pi i}\oint_{C_{z}}
{{\rm d}x\over x-z}A(x)B(z)\,\,\,\,,
\ee
where the countour of integration, $C_{z}$,
surounds the point $z$. This definition
leads to the following form of
Wick's theorem for calculating the product
expansion of $A(z)$ with a composite field $:BC:(w)$
\be
\underbrace{A(z):BC:(w)}={1\over 2\pi i}
\oint_{C_{w}}{{\rm d}x\over (x-w)}
\{\underbrace{A(z)B(x)}C(w)
+(-1)^{BC}\underbrace{A(z)C(w)}B(x)\}\,\,\,,
\ee
where $(-1)^{BC}=-1$ iff both $B$ and $C$ are fermionic fields and
the contraction $(\,\,\underbrace{\,\,\,\,\,\,\,\,\,}\,\,)$
stands for the singular part in the expansion of the product of
two operators at distinct points.
\par
By requiring that the operator product expansion of $T(z)$ with $J^a(z)$
should be given by
\be
T(z)J^a(w)={J^a(w)\over{(z-w)^2}}+
{\partial J^a(w)\over{(z-w)}}
\ee
we get the following equations
\bea
&C_{cb}f^{ba}_{e}+C_{eb}f^{ba}_{c}=0&
\nonumber\\
&D_{\alpha\tau}N^{a\tau}_{\rho}-D_{\rho\tau}N^{a\tau}_{\alpha}=0&
\nonumber\\
&2C_{cb}g^{ba}+C_{bd}f^{ab}_{e}f^{ed}_{c}+D_{\alpha\tau}N^{a\alpha}_\rho
R^{\rho\tau}_c
=\delta^a_c&\,\,\,.
\eea
\par
On the other hand by demanding that the operator product
expansion of $T(z)$ with $S^\alpha(z)$
should be given by
\be
T(z)S^\alpha(w)={S^\alpha(w)\over{(z-w)^2}}+
{\partial S^\alpha(w)\over{(z-w)}}
\ee
leads to
\bea
&D_{\alpha\beta}R^{\beta\gamma}_a + C_{ab}N^{b\gamma}_\alpha=0&
\nonumber\\
&2D_{\alpha\tau}\varphi^{\tau\rho} +
D_{\tau\gamma}R^{\tau\rho}_aN^{a\gamma}_\alpha
+C_{ab}N^{a\rho}_\tau N^{b\tau}_\alpha=\delta^\rho_\alpha&
\eea
\par
The first two equations in (2.14)
and the first equation in (2.16) are equivalent to the set of equations in
(2.8) and are uniquely solved by
\be
C_{ab}=k\Omega_{ab}\,\,\,\,,\,\,\,\,D_{\alpha\beta}=k\psi_{\alpha\beta}\,\,
\,,
\ee
where $k$ is a constant.
Using these solutions in  the last equation  in (2.14), leads to
\be
g^{ab}=-{1\over {2k}}\left[k\left(\gamma^{ab}-\theta^{ab}
\right)-\Omega^{ab}\right]\,\,\,,
\ee
where $\gamma^{ab}$ and $\theta^{ab}$ are defined by
\be
\gamma^{ab}=f^{ea}_cf^{cb}_e\,\,\,,\,\,\,
\theta^{ab}=N^{a\alpha}_\tau N^{b\tau}_\alpha\,\,\,.
\ee
On the other hand, the second equation in (2.16) yields
\be
\varphi^{\alpha\sigma}=-{1\over 2k}\left(2k\Omega_{ab}
N^{a\alpha}_\tau N^{b\tau}_\rho\psi^{\rho\sigma}
-\psi^{\alpha\sigma}\right)\,\,\,.
\ee
A straightforward calculation shows that $g^{ab}$ and
$\varphi^{\alpha\beta}$, as given above, satisfy the relations
dictated by the super Jacobi identities.
\par
With these expressions for $C_{ab}$, $g^{ab}$,
$D_{\alpha\beta}$ and $\varphi^{\alpha\beta}$, the above
energy-momentum tensor satisfies the Virasoro algebra
\be
T(z)T(w)={c\over {2(z-w)^4}}+{2T(w)\over {(z-w)^2}}
+{\partial T(w)\over {(z-w)}}\,\,\,.
\ee
The central charge of this algebra is given by
\be
c=\dim({\cal G}_0) -\dim({\cal G}_1
) - k\Omega_{ab}\left(\gamma^{ab}-3\theta^{ab}
\right)\,\,\,.
\ee
Notice that if $\Omega_{ab}\gamma^{ab}=\Omega_{ab} \theta^{ab}=0$,
then the central charge is given by an integer number equal to
the super dimension of the Lie superalgebra. Before leaving this
section, we would like to mention that equivalent results
have been independently reached in [18].

\section{The WZNW Action for Super Lie Algebras}
An element in the Lie supergroup is experessed as the exponential
\be
g=e^{x_at^a+y_\alpha S^\alpha}\,\,\,,
\ee
where $x_a$ are bosonic variables and $y_\alpha$ are fermionic ones. These
variables are interpreted as coordinates on the group supermanifold.
Notice that $y_\alpha S^\alpha=-S^\alpha y_\alpha$ since they are
fermions.\footnote{If $y_\alpha$ and $S^\alpha$ are matrices then
their product is defined such that $y_\alpha S^\alpha=-S^\alpha y_\alpha$.
See for example [19] for a review on operation on super matrices.}
The WZNW action for Lie superalgebras is constructed out of the
bosonic ``gauge fields" $A_{i a}$
and the fermionic ``gauge fields"
$B_{i \alpha}$, defined for an element $g$ in the
Lie supergroup via
\be
g^{-1}\partial_i g=A_{i a}t^a + B_{i\alpha}S^\alpha\,\,\,.
\ee
Here $i,j,k,\dots$ are world-sheet indices.
\par
Under an infinitesimal transformation of the form
\be
g\rightarrow g + h g + gl\,\,\,,
\ee
where $h$ and $l$ are elements in the  Lie supergroup
and are written as
\bea
h&=&\omega_a t^a + \widetilde\omega_\alpha S^\alpha\nonumber\\
l&=&\theta_a t^a + \widetilde\theta_\alpha S^{\alpha}\,\,\,.
\eea
The variation of the gauge fields is then given by
\bea
A_{i a}&\rightarrow& A_{i a}
+\partial_i\left(\theta_a+\lambda_a\right)
+ A_{i b}\left(\theta_c+\lambda_c\right)
f^{bc}_{a}-B_{i\alpha}\left(\widetilde\theta_\beta
+\widetilde\lambda_\beta\right)R^{\alpha\beta}_a
\nonumber\\
B_{i\alpha}&\rightarrow& B_{i \alpha}
+\partial_i\left(\widetilde\theta_\alpha+\widetilde\lambda_\alpha\right)
+ A_{i a}\left(\widetilde\theta_\beta+\widetilde\lambda_\beta\right)
N^{a\beta}_\alpha-B_{i\beta}\left(\theta_a
+\lambda_a\right)N^{a\beta}_\alpha
\,\,\,,
\eea
with $\lambda_a$ and $\widetilde\lambda_\alpha$ defined by
\be
g^{-1}hg=\lambda_at^a+\widetilde\lambda_\alpha S^{\alpha}\,\,\,.
\ee
An important property of the gauge field $A_{i a}$ and
$B_{i\alpha}$ is that their curvatures vanish
\bea
&{}F_{i j a}=\partial_i A_{j a} -\partial_{j} A_{i a}
+ A_{i c}A_{j d}f^{cd}_{a}
-B_{i\alpha}B_{j\beta}R^{\alpha\beta}_a=0 &
\nonumber\\
&{}H_{i j \alpha}=\partial_i B_{j \alpha} -\partial_{j} B_{i \alpha}
+ A_{i a}B_{j \beta}N^{a\beta}_\alpha
-A_{j a}B_{i \beta}N^{a\beta}_\alpha=0\,\,\,.&
\eea
These last equations will be crucial in demonstrating
the invariance of the action.
\par
Let us also define the following quantities, which will also
enter in determining the Noether currents associated to the
above transformations
\bea
g^{-1}t^ag&=&V^a_bt^b+\widetilde V^a_\alpha S^\alpha
\nonumber\\
g^{-1}S^\alpha g&=&W^\alpha_bt^b+\widetilde W^\alpha_\beta S^\beta\,\,\,.
\eea
Here $(V^a_b,\widetilde W^\alpha_\beta)$
are bosonic quantities while
$(\widetilde V^a_\alpha,W^\alpha_a)$ are fermionic ones.
They satisfy
\bea
\partial_iV^a_b&=&V^a_cA_{id}f^{cd}_b - \widetilde
V^a_\beta B_{i\alpha} R^{\alpha\beta}_b
\nonumber\\
\partial_i\widetilde V^a_\alpha&=&V^a_cB_{i\beta}N^{c\beta}_\alpha
- \widetilde V^a_\beta A_{ic}N^{c\beta}_\alpha
\nonumber\\
\partial_iW^\alpha_a&=&W^\alpha_cA_{id}f^{cd}_a -  \widetilde
W^\alpha_\tau B_{i\beta}R^{\beta\tau}_a
\nonumber\\
\partial_i\widetilde W^\alpha_\beta&=&W^\alpha_cB_{i\tau}N^{c\tau}_\beta
- \widetilde
W^\alpha_\tau A_{ic}N^{c\tau}_\beta\,\,\,.
\eea
These last four equations have been used in proving the invariance
of the action.
\par
Let us now turn our attention to the gauge invariant action.
This is given by
\bea
I(g)&=&-{k\over 8\pi}\int_{\partial B}d^2x\sqrt{-\eta}\eta^{ij}\left(
g^{ab}A_{ia}A_{j b} - \varphi^{\alpha\beta}B_{i\alpha}B_{j\beta}\right)
\nonumber\\
&+&{ik\over 12\pi}\int_{B}d^3y\epsilon^{ijk}\left(g^{ab}
(\partial_i A_{j a})A_{k b}-\varphi^{\alpha\beta}
(\partial_i B_{j \alpha})B_{k \beta}\right)
\,\,\,,
\eea
where $B$ is a three-dimensional manifold
whose boundary is the the two-dimensional surface $\partial B$.
Here we do not require
$g^{ab}$ and $\varphi^{\alpha\beta}$ to be invertible.
The only requirement on these two tensors is that they obey
the relations given by the super Jacobi identities in the
previous section. The above action has already appeared
in [20,21] for non-degenerate $g^{ab}$ and $\varphi^{\alpha\beta}$.
\par
Using the fact that ${}F_{ija}$ and $H_{ij\alpha}$ vanish, and
that $g^{ab}$ and $\varphi^{\alpha\beta}$  are
two invariants of the  Lie superalgebra together with (3.9),
the variation of the action is found to be given by
\bea
&\delta I(g)=-{k\over 4\pi}\int_{\partial B}d^2x
\left(\sqrt{-\eta}\eta^{ij}-i\epsilon^{ij}\right)
\left\{\right.g^{ab}\left[\partial_i\omega_cA_{ja}V^c_b +
\partial_i\widetilde\omega_\alpha A_{ja}W^\alpha_b +
\partial_j\theta_aA_{ib}\right]&
\nonumber\\
&-\varphi^{\alpha\beta}\left[
\partial_i\omega_cB_{j\alpha}\widetilde V^c_\beta
-\partial_i\widetilde\omega_\tau B_{j\alpha}\widetilde W^\tau_\beta
+\partial_j\widetilde\theta_\alpha B_{i\beta}\right]\left.\right\}&
\nonumber\\
&-{ik\over 2\pi}\int_{\partial B}d^2x\epsilon^{ij}
\left[\right.g^{ab}\partial_i\left(\omega_cA_{ja}V^c_b
+ \widetilde\omega_\alpha A_{ja}W^\alpha_b\right)
- \varphi^{\alpha\beta}\partial_i\left(
\omega_cB_{j\alpha}\widetilde V^c_\beta
- \widetilde\omega_\tau B_{j\alpha}\widetilde W^\tau_\beta\right)
\left.\right].&
\eea
Neglecting the last term (surface term), we see that
in complex coordinates $(z,\bar z)$ such that $\eta^{z\bar z}=1$
and $\epsilon^{z\bar z}=i$, the variation of the action
vanishes if $\theta_a=\theta_a(z)$, $\widetilde\theta_\alpha
=\widetilde\theta_\alpha(z)$ and
$\omega_a=\omega_a({\bar z})$, $\widetilde\omega_\alpha=
\widetilde\omega_\alpha(\bar z)$. The Noether currents
associated to some  Lie
superalgebra elements,  $t^a$ and $S^\alpha$ are given by
\bea
&J^a_z=-{k \over 2\pi}g^{ab}A_{zb}\,\,\,\,,\,\,\,\,
J^a_{\bar z}=-{k \over 2\pi}g^{bc}A_{\bar z b}V^a_c
+{k\over 2\pi}\varphi^{\alpha\beta}B_{\bar z \alpha}\widetilde
V_\beta^a&
\nonumber\\
&S^\alpha_z={k\over 2\pi}\varphi^{\alpha\beta}B_{ z \beta}
\,\,\,\,,\,\,\,\,
S^\alpha_{\bar z}=-{k \over 2\pi}g^{bc}A_{\bar z b}W^\alpha_c
-{k\over 2\pi}\varphi^{\tau\beta}B_{\bar z \tau}\widetilde
W_\beta^\alpha&
\eea
The sets of currents $(J^a_z\,,\,S^\alpha_z)$
and $(J^a_{\bar z}\,,\,S^\alpha_{\bar z})$ are, by virtue
of the equations of motion, holomorphic and antiholomorphic
currents, respectively. In the case when $g^{ab}$ and
$\varphi^{ab}$ are invertible, these currents satisfy
two commuting copies of the supercurrent algebra
given in (2.7).
\par
In summary, we have considered in this paper current algebras
based on Lie superalgebras.  We gave a systematic approach
to constructing the energy-momentum tensor with respect
to which the supercurrents are primary fields of conformal dimension
one.
The Virasoro central charge for certain cases is an integer number
equal to the super dimension of the  Lie superalgebra.
A Wess-Zumino-Novikov-Witten action based on Lie
superalgebras is also constructed.
\par
The study carried out in this paper raises two questions.
The first one is whether the WZNW action for Lie
superalgebras would define some string backgrounds. In order to answer
this question on has to determine the conformal invariance conditions
(the beta functions) for a non-linear sigma model defined on a
supermanifold. The second question concerns the connection between
our construction and topological field theory where  Lie superalgebras are
of crucial importance [21,22]. It would also be very desirable to
generalise the theorems of ref.[6] to Lie superalgebras. Work in this
direction
is already in progress in the first reference of [18].
\vspace{0.5cm}
\paragraph{Acknowledgements:}
I would like to thank Alex Deckmyn for motivating
me to look into this problem. It is also a pleasure to thank
Jos\'e M. {}Figueroa-O'{}Farrill,
Werner Nahm and Manfred Scheunert for  many useful discussions.
This research is  supported  by the Alexander von Humboldt-Stiftung.

\end{document}